\documentclass[twocolumn]{aastex62}

\received{January 15, 2019}
\accepted{February 20, 2019}
\submitjournal{AJ}

\shorttitle{Orbital period variation of KIC\,10544976}
\shortauthors{Almeida et al.}

\begin{document}

\title{\Large \bf Orbital period variation of KIC\,10544976: Applegate mechanism versus light travel time effect}

\correspondingauthor{Leonardo~A. Almeida}
\email{leonardoalmeida@fisica.ufrn.br}

\author[0000-0002-3817-6402]{Leonardo~A. Almeida}
\affil{Departamento de F\'isica Te\'orica e Experimental, Universidade Federal do Rio Grande do Norte, CP 1641, Natal, RN, 59072-970, Brazil}
\affil{Instituto de Astronomia, Geof\'isica e Ci\^encias Atmosf\'ericas, Rua do Mat\~ao 1226, Cidade Universit\'aria S\~ao Paulo, SP, Brasil, 05508-090}

\author{Leandro de Almeida}
\affil{Departamento de F\'isica Te\'orica e Experimental, Universidade Federal do Rio Grande do Norte, CP 1641, Natal, RN, 59072-970, Brazil}

\author{Augusto Damineli}
\affil{Instituto de Astronomia, Geof\'isica e Ci\^encias Atmosf\'ericas, Rua do Mat\~ao 1226, Cidade Universit\'aria S\~ao Paulo, SP, Brasil, 05508-090}

\author{Claudia V. Rodrigues} 
\affil{Instituto Nacional de Pesquisas Espaciais/MCTIC, Avenida dos Astronautas 1758, S\~ao Jos\'e dos Campos, SP, 12227-010, Brazil} 

\author{Matthieu Castro}
\affil{Departamento de F\'isica Te\'orica e Experimental, Universidade Federal do Rio Grande do Norte, CP 1641, Natal, RN, 59072-970, Brazil}

\author{Carlos E. F. Lopes}
\affil{Instituto Nacional de Pesquisas Espaciais/MCTIC, Avenida dos Astronautas 1758, S\~ao Jos\'e dos Campos, SP, 12227-010, Brazil}

\author{Francisco Jablonski}
\affil{Instituto Nacional de Pesquisas Espaciais/MCTIC, Avenida dos Astronautas 1758, S\~ao Jos\'e dos Campos, SP, 12227-010, Brazil}

\author{Jos\'e D. do Nascimento Jr.}
\affil{Departamento de F\'isica Te\'orica e Experimental, Universidade Federal do Rio Grande do Norte, CP 1641, Natal, RN, 59072-970, Brazil}
\affil{Harvard-Smithsonian Center for Astrophysics, 60 Garden St., Cambridge, MA 02138, USA}

\author{Marildo G. Pereira}
\affil{Universidade Estadual de Feira de Santana, Av. Transnordestina, S/N, Feira de Santana, BA, 44036-900, Brazil}

\begin{abstract}

In recent years, several close post-common-envelope eclipsing binaries have been found to show cyclic eclipse timing variations (ETVs). This effect is usually interpreted either as the gravitational interaction among circumbinary bodies and the host binary -- known as the light travel time (LTT) effect -- or as the quadrupole moment variations in one magnetic active component -- known as Applegate mechanism. In this study, we present an analysis of the ETV and the magnetic cycle of the close binary KIC\,10544976. This system is composed of a white dwarf and a red dwarf in a short orbital period (0.35~days) and was monitored by ground-based telescopes between 2005 and 2017 and by the {\it Kepler} satellite between 2009 and 2013. Using the {\it Kepler} data, we derived the magnetic cycle of the red dwarf by two ways: the rate and energy of flares and the variability due to spots. Both methods resulted in a cycle of $\sim$600~days, which is in agreement with magnetic cycles measured for single low-mass stars. The orbital period of KIC\,10544976 shows only one long-term variation which can be fitted by an LTT effect with period of $\sim$16.8~yr. Hence, one possible explanation for the ETVs is the presence of a circumbinary body with minimal mass of $\sim$13.4~$M_{\rm Jup}$. In the particular scenario of coplanarity between the external body and the inner binary, the third body mass is also $\sim$13.4~$M_{\rm Jup}$. In this case, the circumbinary planet must either have survived the evolution of the host binary or have been formed as a consequence of its evolution.

\end{abstract}


\section{INTRODUCTION}\label{sect:intro}

Eclipse timing variation (ETV) is a common phenomenon in compact evolved binaries. These variations have been observed in HW Vir systems \citep[e.g.][]{Lee+2009,Almeida+2013}, DA+dM eclipsing binaries \citep[e.g.][]{Parsons+2010,Beuermann+2011,Qian+2012}, cataclysmic variables (e.g. \citealt{Dai+2010}), and RS CVn binaries \citep[e.g.][]{Liao+2010}. A review on the orbital period variation of evolved compact binaries is found in \citet[][]{Zorotovic+2013}.

\begin{figure*}
\centering
 \begin{minipage}{180mm}
 \resizebox{\hsize}{!}{\includegraphics{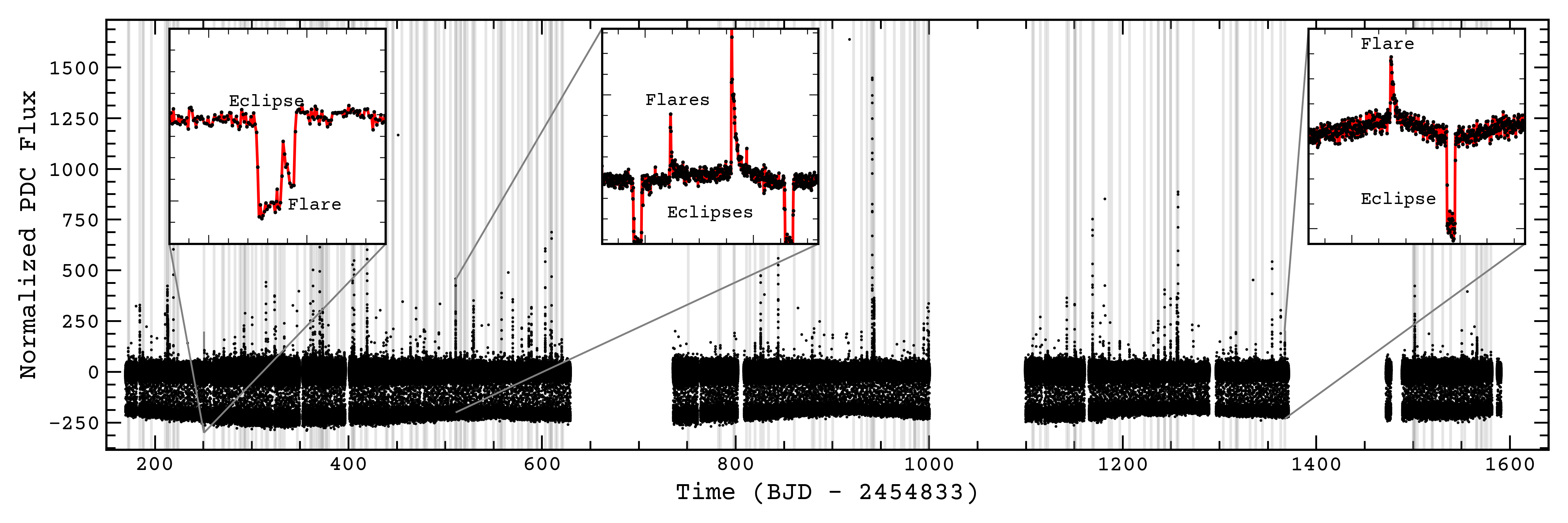}}
 \caption{Normalized PDC short-cadence light curves of KIC\,10544976 obtained by the {\it Kepler} satellite. The gray vertical lines indicate the flare locations. Three parts of the light curve are zoomed to better visualize the flares, the reflection effect, and a flare inside an eclipse.}
 \label{lightcurves}
 \end{minipage}
\end{figure*}

Changes in the orbital period of detached evolved compact binaries in the timescale of years to decades have been explained either by additional components interacting gravitationally with the binary causing the so-called light travel time (LTT) effect \citep[e.g.][]{Lee+2009,Beuermann+2011,Qian+2012} or by magnetic activity cycles (MACs) of the main-sequence components, also known as the Applegate mechanism \citep[e.g.][]{Applegate1992,Parsons+2010,Bours+2016}. The cause of ETV is still an open issue in the literature \cite[e.g.][]{Bonavita+2016,Bours+2016,Volschow+2016,Volschow+2018,Pulley+2018}. One possible way to disentangle the problem is to look for independent evidence of third bodies or MACs. The results on the presence of third bodies are so far inconclusive \cite[see e.g.][]{Marsh+2014,Hardy+2015,Hardy+2016,Marchioni+2018}. Additionally, information about the MACs in systems presenting ETV is still lacking.

KIC\,10544976 (J2000: RA $=$ 19$^{\rm h}$42$^{\rm m}$37${\fs}$2, decl. $=$ +47$^{\circ}$45$^\prime$48$\farcs$7) is an eclipsing post-common-envelope binary consisting of a white dwarf (WD; primary) and a magnetically active M3.5-M4 dwarf star (dM; secondary) in a close orbit, $P_{\rm orb}\sim0.35$~day. The component masses are $M_{\rm 1} = 0.61\, M_{\odot}$ and $M_{\rm 2} = 0.39\, M_{\odot}$, totaling 1~$M_{\odot}$ \citep[][]{Almenara+2012}. This object was observed by the {\it Kepler} satellite for $\sim$3.9 yr in the short- and long-cadence modes and has been monitored since 2005 from ground-based telescopes. Searches for a third body around this system using only {\it Kepler} data were not successful \citep[e.g.,][]{Conroy+2014AJ,Gies+2015,Borkovits+2016}. However, this system has almost 12 yr of observational coverage and a significant number of flares in the {\it Kepler} light curve. Thus, for the first time, the same object has suitable data to measure the ETV and its MAC, allowing us to test whether these effects are correlated.

In this study, we present an orbital period analysis of KIC~10544976 and explore the two main possible scenarios that could explain the results. In Section~\ref{observation}, we describe the {\it Kepler} data and the data reduction of the William Herschel and Isaac Newton telescopes (INT). Section~\ref{analysis} presents the methodology used to obtain the mid-eclipse times, the procedures to examine the orbital period variation in the context of the LTT effect, and our approach to determine the MAC. Finally, we discuss the results in Section~\ref{discussion}.

\section{OBSERVATIONS AND DATA REDUCTION}\label{observation}

\subsection{Kepler Satellite}

The KIC\,10544976 data were retrieved from the Kepler data search webpages\footnote{\tt http://exoplanetarchive.ipac.caltech.edu}$^{,}$\footnote{\tt https://archive.stsci.edu/pub/kepler/ffi}. We use the Presearch Data Conditioning\footnote{\tt https://keplergo.arc.nasa.gov/PipelinePDC.shtml} (PDC) and the full frame images (FFIs). The PDC extracts the flux via aperture photometry and corrects the systematic effects, e.g., outliers and discontinuities within the quarters \citep[e.g.][]{Jenkins+2010}. The KIC\,10544976 light curve is composed by short and long cadences. We use both cadences, which have integration time of $\sim$1 and $\sim$30~minutes covering $\sim$3.9~yr from 2009 to 2013. Figure~\ref{lightcurves} shows the normalized short-cadence light curve (as described in Sec. \ref{sec:flare}).

The FFIs are images of the {\it Kepler}'s entire field of view. There are 53 FFIs, of which 8 of them were collected in the 34 hr period during the commissioning of the spacecraft in 2009, and the other images were obtained approximately one per month throughout the primary mission. The FFIs have the same integration time and calibration of the long-cadence data.

\subsection{Isaac Newton Group of Telescopes}

Raw photometric data of KIC\,10544976 and calibration (flat-field and bias) images were retrieved from the Isaac Newton Group public archive\footnote{\tt http://casu.ast.cam.ac.uk/casuadc/ingarch/query}. The photometric observations were performed with the Auxiliary-port CAMera attached to the $4.2$ m William Herschel Telescope (WHT) and with the Wide Field Camera coupled to the $2.54$ m INT at Roque de los Muchachos Observatory. The observations were done on 2008 August 28 and 2014 July 21 (WHT), and on 2017 April 05  (INT). These data comprise 344 images ($t_{\rm exp}$ = 10~s) in the V band and 1209 images ($t_{\rm exp}$ = 5~s) plus 136 images ($t_{\rm exp}$ = 25~s) unfiltered.

The WHT and INT data were reduced using the standard IRAF\footnote{\tt http://iraf.noao.edu} tasks, which subtract a master median bias image from each program image, divide the result by a normalized flat field, and perform differential photometry. As the KIC\,10544976 field is not crowded, aperture photometry was used to extract its flux relative to a constant target (2MASS\,19423960+4744583). 

\section{ANALYSIS AND RESULTS}\label{analysis}

\subsection{Eclipse fitting}
\label{eclipe_fit}

The mid-eclipse times of KIC\,10544976 were obtained performing a model fit to each observed eclipse using the Wilson-Devinney code (WDC; \citealt{Wilson+1971}). The range of the geometrical and physical parameters, e.g., inclination, radii, and temperatures obtained by \citet[][]{Almenara+2012}, were adopted as search intervals in the fitting procedure. 

We use the same approach described in \citet{Almeida+2013} to derive the mid-eclipse times and their uncertainties. In short, a Markov chain Monte Carlo (MCMC) procedure was performed to sample the mid-eclipse time and to examine the marginal posterior distribution of the probability of the parameters. The median of the distribution provides the mid-eclipse time and the area corresponding to the 1$\sigma$ percentile provides the uncertainty. Our 2745 new mid-eclipse times from {\it Kepler}, WHT and INT data, and previous values obtained by \citet{Almenara+2012} are presented in Table~\ref{tab:mid}. We reanalyzed the data collected on 2008 August 28 presented in \citet{Almenara+2012}. Our results are in agreement. 

\begin{table}
\centering
\caption{Mid-eclipse times of KIC~10544976.} 
\label{tab:mid}  
\begin{tabular}{l c c r }        
\hline
Cycle  & Time  &  Error  & Reference   \\ 
       & (BJD-245,000) & (d)           &                   \\ 
\hline                        
1    &  3590.43688       & 0.00002 & \citealt{Almenara+2012}   \\
4    &  3591.48829       & 0.00001 &  \citealt{Almenara+2012}   \\
7    &  3592.53970       & 0.00005 &  \citealt{Almenara+2012}   \\
10   &  3593.59112       & 0.00005 &  \citealt{Almenara+2012}   \\
13   &  3594.64256       & 0.00007 &  \citealt{Almenara+2012}   \\
21   &  3597.44625       & 0.00003 &  \citealt{Almenara+2012}   \\
1011 &  3944.41025       & 0.00003 &  \citealt{Almenara+2012}   \\
1014 &  3945.46169       & 0.00007 &  \citealt{Almenara+2012}   \\
1017 &  3946.51304       & 0.00002 &  \citealt{Almenara+2012}   \\
3143 &  4691.60964       & 0.00004 &  This work       \\
\hline         
\end{tabular}%
\tablecomments{Table 1 is published in its entirety in the machine-readable format.}
\end{table}

\subsection{Orbital Period Variation}

To analyze the ETV of KIC\,10544976, we initially examined if its orbital period could be represented by a linear ephemeris as

\begin{equation}
 T_{\rm min} ({\rm TBD})= T_0 + E\times P_{\rm bin},
 \label{lin_efem}
\end{equation}

\noindent where $T_{\rm min}$ are the mid-eclipse times, $T_0$ is an initial epoch, $E$ is the cycle count from $T_0$, and $P_{\rm bin}$ is the orbital period. The residuals of the mid-eclipse times with respect to a linear ephemeris, also known as the observed minus calculated ($O-C$) diagram, show an apparent cyclic variation (see top panel of Figure~\ref{figo-c}). In the following sections, we verify if the ETV can be explained by the Applegate mechanism or by the LTT effect.

\subsubsection{Light Travel Time Effect}\label{section:ltt}

The LTT effect in the received times of a periodic source is generated by variations in the distance between this source and the observer. In the case of an eclipsing binary, the eclipse times are modulated as a result of the gravitational interaction of the binary with other bodies \citep{irwin1952}. Adding this effect to Eq.~\ref{lin_efem}, we obtain

\begin{equation}
T_{\rm min} = T_0 + E\times P_{\rm bin} + \sum_1^n \tau_{\rm j},
\label{ephem_ltt}
\end{equation}

\noindent where

\begin{equation}
\tau_{\rm j} = \frac{z_{\rm j}}{c}=a_{\rm bin; j}\sin i_{\rm j}\left[\frac{1-e^2_{\rm j}}{1+e_{\rm j} 
\cos f_{\rm j}} \sin (f_{\rm j}+\omega_{\rm j})\right]
\label{ltt}
\end{equation}

\noindent is the LTT effect of the $j$th body, $c$ is the speed of light, $a_{\rm bin;j}$ is the semi-major axis, $e_{\rm j}$ is the eccentricity, $i_{\rm j}$ is the inclination, $f_{\rm j}$ is the true anomaly, and $\omega_{\rm j}$ is the argument of periastron. 

\begin{figure}
\centering
\includegraphics[angle=0,scale=0.42]{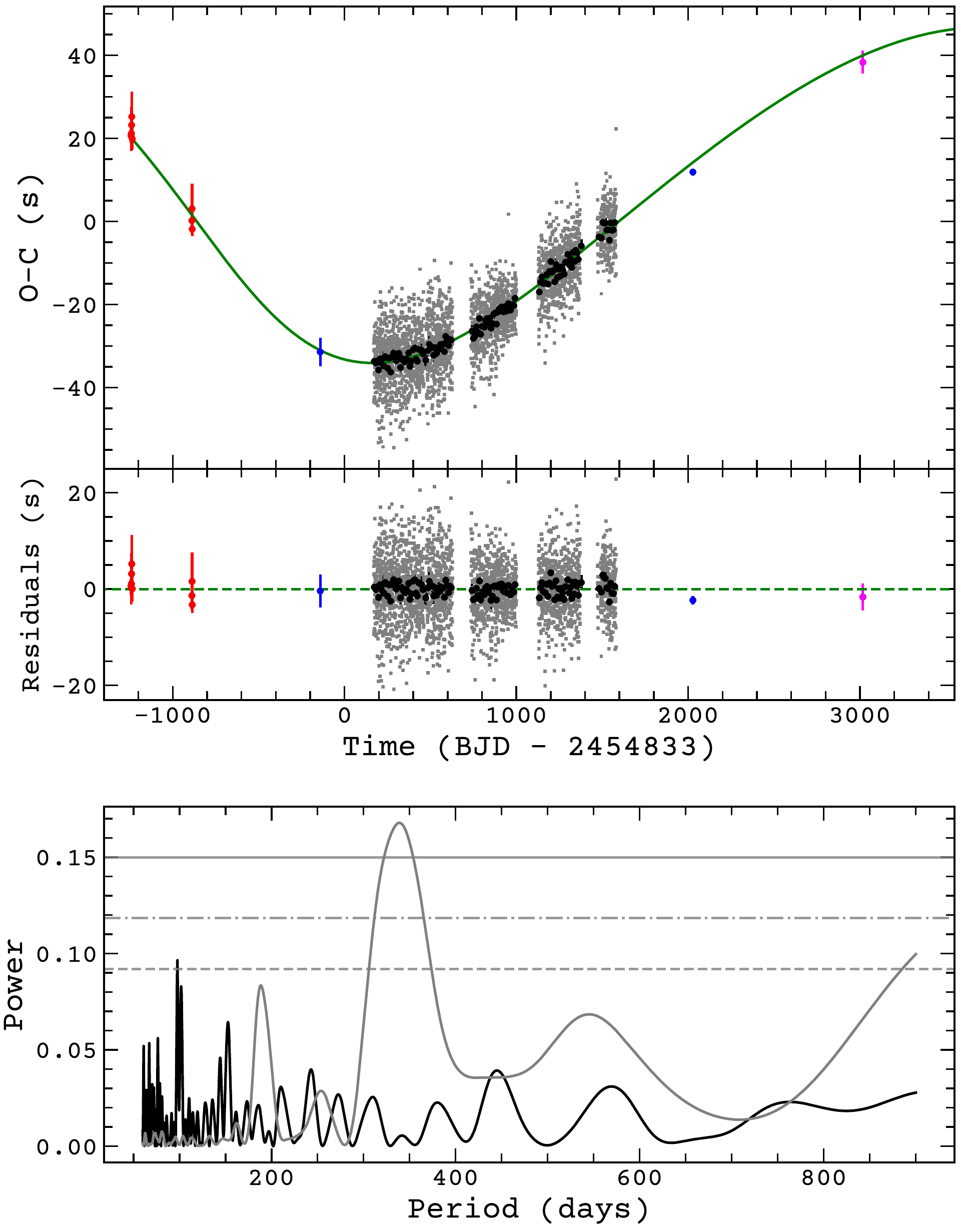}
\caption{Top panel: $(O-C)$ diagram of KIC\,10544976. Individual and average of 20 mid-eclipse times of the {\it Kepler} data are shown with gray and black points, respectively. Measurements from \citet{Almenara+2012}, WHT, and INT data are shown in red, blue, and magenta points. The solid line represents the best fit including one LTT effect. Middle panel: the residuals around the best fit. Bottom panel: Lomb--Scargle periodograms of the residuals (black line) and their spectral window (gray line). Solid, dashed-dotted, and dashed horizontal lines represent the false alarm probability (FAP) for 1\%, 10\%, and 50\%, respectively.}
\label{figo-c}
\end{figure}

Using Equation~\ref{ephem_ltt} with only one LTT effect to fit the mid-eclipse times, the resulting $\chi_{\rm red}^2$ is 1.5. In the fitting procedure, we use the \textsc{Pikaia} algorithm \citep[][]{Charbonneau1995} to look for the global solution, followed by a MCMC to sample the parameters of Eq.~\ref{ephem_ltt} around this solution. Figure~\ref{figo-c} shows graphically the results and Table~\ref{tab:pars} shows the fitted and derived parameters.

\subsubsection{Applegate Mechanism}

The Applegate mechanism \citep{Applegate1992} consists of the coupling between the binary orbit and changes in the shape of a magnetically active component. These changes are generated by variations in the quadrupole moment, which are directly related to the MAC and should lead to cyclic variations of the binary orbital period.

\citet{Brinkworth+2006} expanded the Applegate original expressions to include a stellar, thick outer shell. They showed that the ETV of NN~Ser cannot be explained by the Applegate mechanism based on energetic grounds. Recently, \citet{Volschow+2016} extended the formulation developed by \citet{Brinkworth+2006} including quadrupole moment changes in two finite regions, the core and the external shell. They applied this model for 16 compact binaries and concluded that the Applegate mechanism can explain the ETV for 4 systems. 

In the Applegate mechanism scenario, the observed ETV should have the same period as the MAC. In the following sections, we present the procedure to derive the MAC of KIC\,10544976 via the flare occurrence rate and modulations by stellar spots.

\begin{table}
\caption{Parameters for the linear ephemeris plus one LTT effect of KIC\,10544976.} 
\centering 
\label{tab:pars}  
\begin{tabular}{l c r }       
\hline            
 & \small Linear Ephemeris  &  \\
\hline 
Parameter  & Value & Unit   \\   
\hline                       
$P$        & $0.350468924\pm 1 \times 10^{-9}$   & d  \\
$T_0$      & $2453590.08616\pm 5 \times 10^{-5}$ & BJD(TDB)  \\
\hline
  & \small $\tau_1$ term   &  \\
\hline
Parameter  & Value & Unit   \\    
\hline                        
$P$           &  $16.8\pm2.4$                 & yr  \\
$T$           & $2454254\pm244$               & BJD(TDB)   \\
$a_{\rm bin}\sin i$& $0.084\pm0.014$          & au     \\
$e$           & $0.29\pm0.11$                 &        \\
$\omega$      & $211\pm17$                   & $^\circ$ \\
$f(m)$        & $(2.1\pm1.2) \times 10^{-6}$  & $M_{\odot}$  \\ 
$m_{\rm min}$ & $13.4\pm1.0$               & $M_{\rm Jup}$  \\  
$a_{\rm min}$ & $6.56\pm1.74$             & au \\ 
\hline
\end{tabular}
\end{table}

\subsubsection{Flare Analysis}\label{sec:flare}

The {\it Kepler} light curve of KIC\,10544976 presents many flares. The flares are observed in any phase of the orbital cycle, including in the primary eclipses, which points to an origin in the secondary star, probably due to its magnetic activity (see Figure~\ref{lightcurves}). The occurrence rate and energy of the flares have been used to derive the stellar magnetic cycle. The periodicity of these quantities are well known to be in good agreement with the MAC derived via photometric modulations generated by stellar spots for the Sun, solar-type, and low-mass stars \citep[e.g.][]{Basri+2010,Savanov2012,Mathur+2014,Montet+2017ApJ,He+2018}. 

To calculate the frequency and energy of the KIC\,1054\-4976 flares in the short-cadence {\it Kepler} light curve, we use the following steps.

\noindent (i) The jumps in the PDC light curves among the quarters were corrected and then, the whole data normalized. 

\noindent (ii) A phase diagram was built using the orbital period and initial epoch shown in Table~\ref{tab:pars} in order to subtract the reflection effect by fitting a sinusoidal function and to shift the inner part of the WD eclipses to the external level. 

\noindent (iii) An automatic task was developed to find the flares. Our task searches for outliers (or peaks), whose we consider as flare candidates when three consecutive points follow the criterion

\begin{equation}
\frac{\left( y_{i} - y_{mL}\right)}{\sigma_{mL}} \geq 3,
\label{eq_sel}
\end{equation}   

\noindent where $y_{mL}$ and $\sigma_{mL}$ are the average and the standard deviation computed in the box of fixed $L = 0.035$~d size. The false flare candidates, those that do not follow the flare standard behavior, i.e., a rapid rise and a slow exponential decrease, were removed by visual inspection and $211$ flare candidates were confirmed as bona fide flares. The positions of all the flares in the light curve are shown in Figure~\ref{lightcurves} with yhe gray vertical lines. To search for periodicity in the flare occurrence rate, we count the flares in boxes with the same timescale as the files available by {\it Kepler}, which have $\sim$30 days. The number of flares, their Lomb--Scargle (LS) periodogram, and the phase diagram for the frequency of maximum power (equivalent to $P = 600\pm134$~days) are shown in the top panels of Figure~\ref{fig:nrp_index} (yellow marks). The frequency uncertainty was derived by fitting a Gaussian function in the maximum peak of the periodogram.

\noindent (iv) We also use a flare index $P_{\rm flare}$ proposed by \citet[][]{He+2018}. This index is equivalent to the flare energy rate. The results for $P_{\rm flare}$ taking the same size boxes as used for the flare counting is shown in the top left panel of Figure~\ref{fig:nrp_index} (blue marks). To search for periodicity, we removed a point that has more than 3$\sigma$ above the average of the whole sample. This point (shown as an empty square in the $P_{\rm flare}$ data in Figure~\ref{fig:nrp_index}) is due to a long-lasting flare and has a strong influence in the search for periodicity based in the LS method. The LS periodogram and phase diagram are displayed in the top right and middle panels of Figure~\ref{fig:nrp_index}. The maximum power frequency is equal to $550\pm101$~days.

\subsubsection{Spot Modulation Analysis}

As the secondary star shows a significant number of flares, a contribution of the spots is expected in the Kepler light curve. This effect can also provides information about magnetic cycles \citep[e.g.][]{Olah+2009,Savanov2012,Vida+2013,Vida+2014,FerreiraLopes+2015,Montet+2017ApJ}. This question was addressed analyzing the PDC long-cadence light curve and the FFIs in the following way. 

\noindent(i) The PDC long-cadence light curve was normalized following the same procedure used for the flare analysis, see Section~\ref{sec:flare}. 

\noindent(ii) We applied multiplicative factors among the quarters to account for the different aperture masks adopted by the {\it Kepler} pipeline. These factors were calculated to produce equal eclipse depths and reflection effect amplitudes among the quarters. We are confident that this procedure preserves the original long-term behavior of the KIC\,10544976 light curve because the signal of $\sim$372.5 days due to the {\it Kepler} heliocentric orbital period was conserved \citep[see, e.g.,][]{VanEylen+2013}. 

\noindent(iii) The flare and eclipse regions were cut and a sigma clipping (3$\sigma$) was performed to remove other outliers. The 372.5 day signal was also subtracted using a sinusoidal fit. 

\begin{figure*}
\centering
 \begin{minipage}{180mm}
\resizebox{\hsize}{!}{\includegraphics{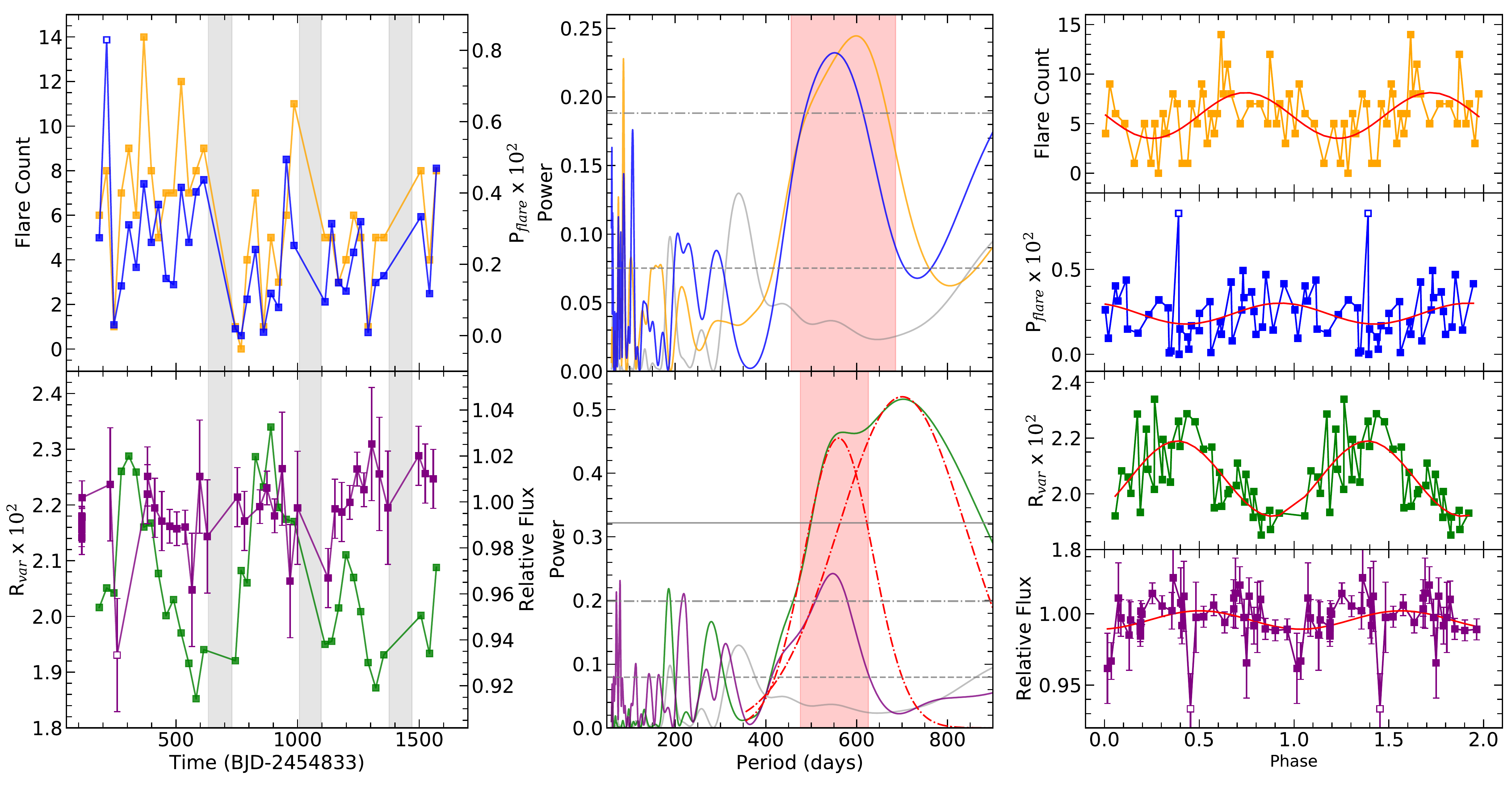}}
\caption{Top panels: from left to right, flare count (yellow marks) and flare energy rate ($P_{\rm flare}$; blue marks), their LS periodograms, and phase diagrams for the maximum power frequencies. Left panel: gray areas are the temporal ranges without data. Middle panel: the solid gray line represents the LS periodogram for the spectral window, the dashed and dashed--dotted lines represent the FAP for 50\% and 10\%, respectively. The pink region displays the $\pm 1\,\sigma$ uncertainty around the maximum peak of the flare count. Top and bottom right panels: red-solid lines exhibit the sinusoidal fits keeping the periods corresponding to the maximum power frequencies of the flare count and $P_{\rm flare}$ fixed. Bottom panels: same as the top panels but for the activity proxy ($R_{\rm var}$; green marks) and relative flux (magenta marks), with exception that, in the middle panel the maximum power in the $R_{\rm var}$ LS periodogram has two peaks which were fitted with two Gaussian functions (dashed--dotted red lines), the horizontal solid line is the FAP for 1\%, and the pink area is $\pm 1\,\sigma$ uncertainty around the maximum peak of the relative flux.}
 \label{fig:nrp_index}
 \end{minipage}
\end{figure*}

\begin{figure}
\includegraphics[angle=0,scale=0.37]{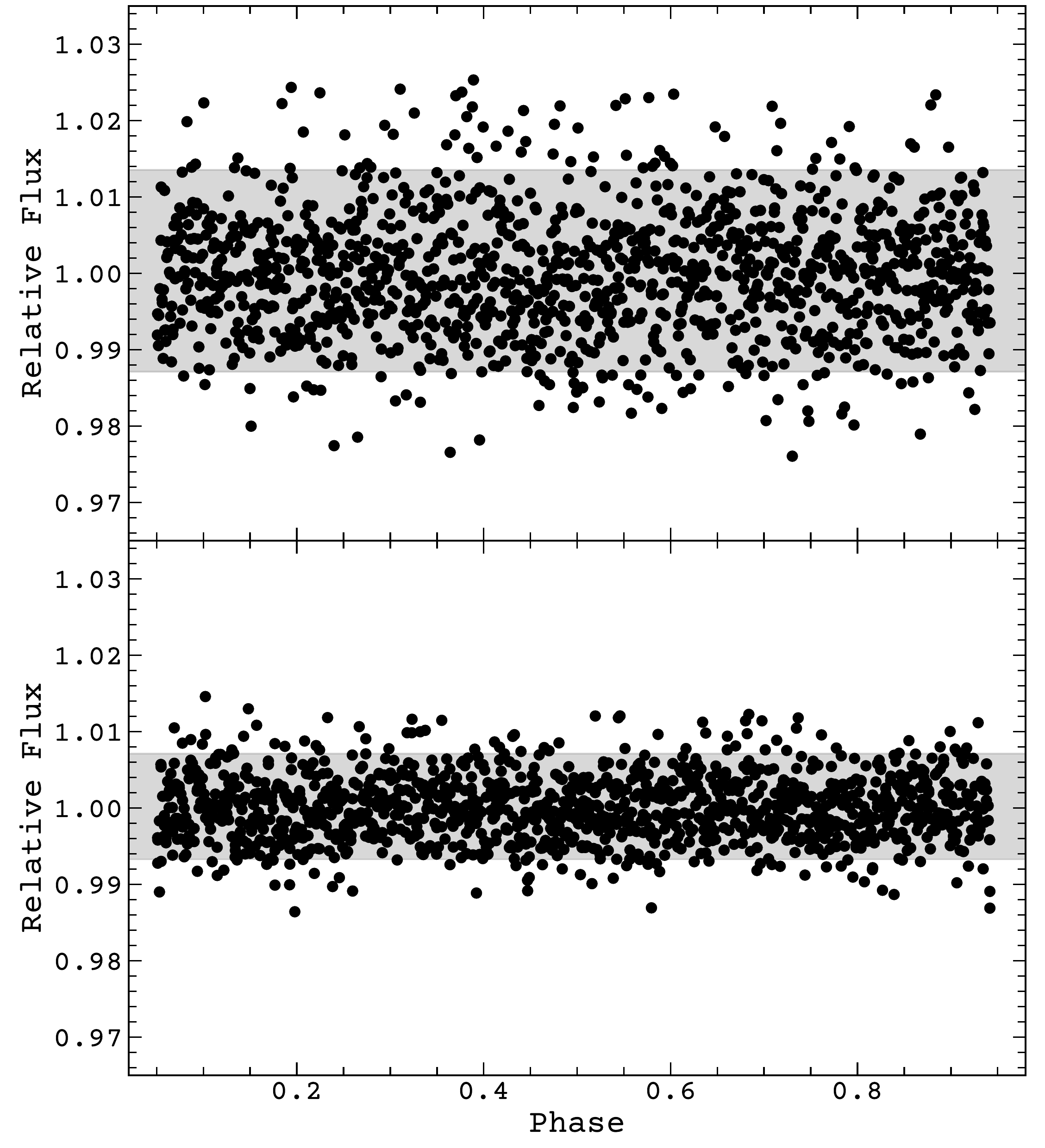}
\caption{Normalized PDC long-cadence light curve of KIC\,10544976 with high and low incidence of spots. Top panel: phased light curve with a time range of 30 days around the day 306 in the {\it Kepler}'s reference time. The gray area displays the region between 5\% and 95\% percentile intensities. Bottom panel: same as the upper panel around the day 583 in the {\it Kepler}'s reference time.}
 \label{fig:rvar}
\end{figure}

\noindent(iv) The stellar activity proxy $R_{\rm var}$, defined as the difference between the 5\% and 95\% percentile intensities \citep[][]{Basri+2010}, was computed in time boxes as those used for the flares. The results for $R_{\rm var}$ are present in the bottom panels of Figure~\ref{fig:nrp_index} (green marks). The maximum power frequency in the LS periodogram show two peaks, $585\pm92$ and $694\pm130$~days. As an example of how the spots increase the dispersion in the light curve, Figure~\ref{fig:rvar} shows two normalized and phased light curves of 30 day ranges, which have high (top panel) and low (bottom panel) incidence of spots.

\noindent(v) We applied the method developed by \citet[][]{Montet+2017ApJ} to measure the long-term variability of KIC\,10544976 using the FFIs. Their Full Frame Fotometry package\footnote{https://github.com/benmontet/f3} was used to obtain the relative flux of our target with respect to nearby reference stars on the detector. To correct the variability due to the binary variation, the first eight images, obtained during 34 hr, were modeled using the WDC, and this information was applied to all other 32 available FFIs for KIC\,10544976. The result of this procedure, its LS periodogram, and the phase diagram for the period of $544\pm88$~days, which corresponds to the highest power frequency, are presented in the  bottom panels of Figure~\ref{fig:nrp_index} (magenta marks). One point was removed in the search for periodicity using the same criterion adopted for the $P_{\rm flare}$ (see the empty square in the bottom panel of Figure~\ref{fig:nrp_index}).

\section{DISCUSSION AND CONCLUSIONS}
\label{discussion}

In this study, we present an orbital period variation analysis of the post-common-envelope binary KIC\,10544976. We combined 2745 new mid-eclipse times from {\it Kepler}, WHT, and INT data with measurements available in the literature and performed a linear ephemeris fitting. One cycle with a semi-amplitude of $\sim$35~s is clearly seen in the $(O-C)$ diagram with a period of $\sim$17~yr. We explore the two possible scenarios, the LTT effect and Applegate mechanism, that could explain the cyclic changes in the orbital period of such systems in timescales from years to decades. 

Assuming the LTT effect scenario, the best solution yields an orbital period of  $16.83\pm2.4$~yr, eccentricity of $0.29\pm0.11$,  and minimal mass of $\sim$13.4~$M_{\rm Jup}$ for the external body. In the particular scenario of coplanarity between the external body and the host binary, knowing the mass and inclination of the inner binary, $1.0~M_{\odot}$ and 89$\fdg$6 \citep{Almenara+2012}, the mass of the circumbinary body would be $\sim$13.4~$M_{\rm Jup}$. In this case, KIC\,10544976 would be composed by an inner binary and a giant planet. However, as the third body orbital period is longer than the observational baseline, this solution must be considered as a preliminary one. Therefore, new eclipse time measurements are essential to confirm our solution.

In the Applegate mechanism context, what has been done for several authors \citep[see, e.g.,][]{Brinkworth+2006,Qian+2012,Volschow+2016} is to verify if the secondary star has the required energy to produce the observed variation amplitude in the $(O-C)$ diagram. Following \citet{Volschow+2016} and using their online calculator\footnote{http://theory-starformation-group.cl/applegate/index.php}, we obtained that the required energy is much larger than the energy of the secondary star. Therefore, the Applegate mechanism cannot explain the ETV of KIC\,10544976. However, for the first time, we are able to perform another test. As this system harbors an active secondary star and has almost 3.9 yr of {\it Kepler} data, it is possible to search for the magnetic cycle using the flare counting and energy rate ($P_{\rm flare}$) as well as modulations and long-term variations in the light curve generated by stellar spots. The LS periodograms of the flare count and $P_{\rm flare}$ show maximum peaks at $600\pm134$~days and $550\pm101$ days, which agree within the uncertainties. Whereas for the spot modulations, represented by the stellar activity proxy $R_{\rm var}$, the maximum frequency in the LS periodogram has two peaks ($585\pm92$ and $694\pm130$~days), the long-term variability, expressed by the relative flux, shows only one with a period of $544\pm88$~days. Hence, all periods associated with flares and stellar spots agree with each other within the uncertainties, see Figure~\ref{fig:nrp_index}. Therefore, we conclude that the MAC of the KIC\,10544976 secondary star has a period around 600 days. 

\begin{figure}
\includegraphics[angle=0,scale=0.42]{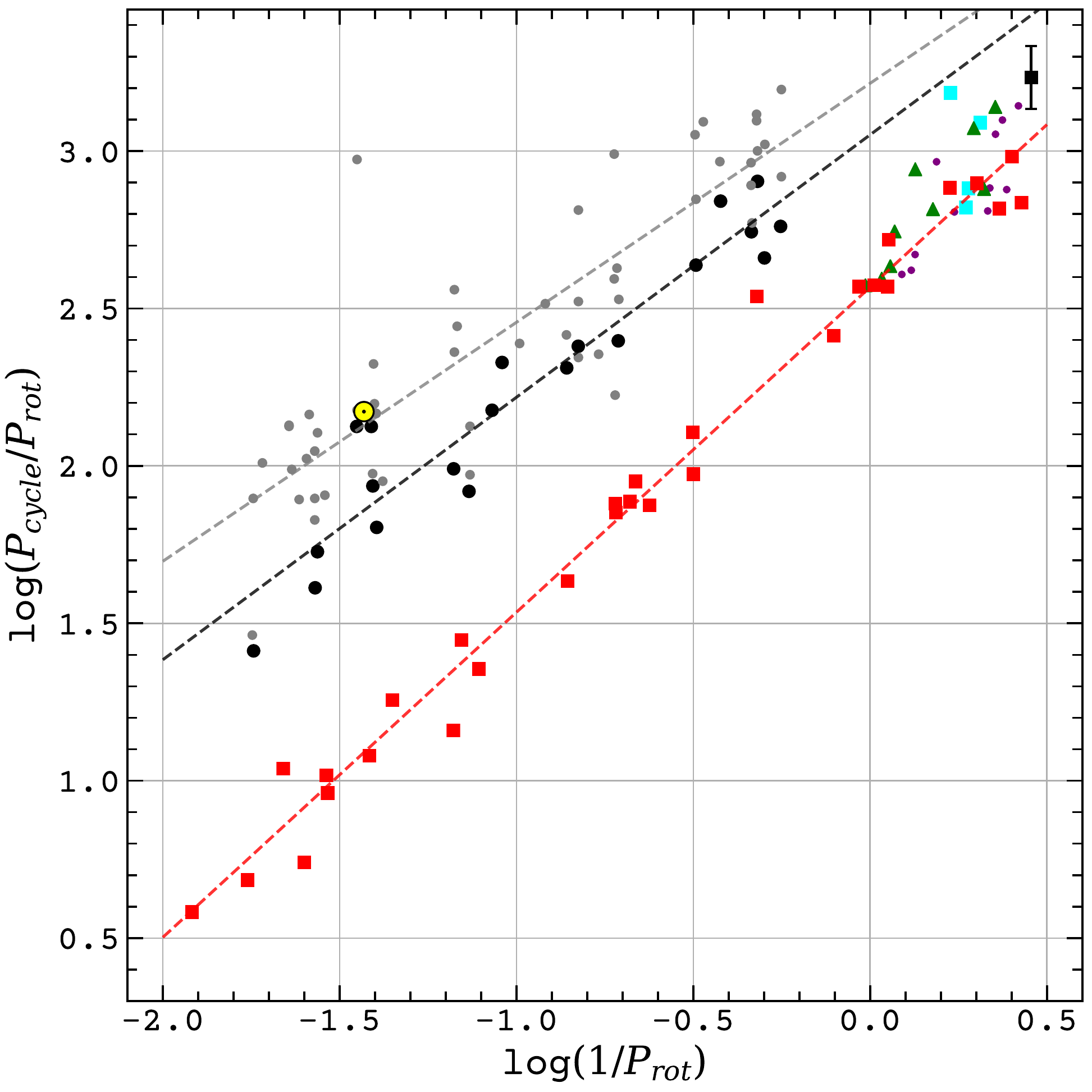}
\caption{Diagram relating the MAC and the rotational period as shown in \citet[][]{Vida+2014}. The black dots, blue squares, and green triangles represent measurements from \citet[][]{Vida+2013,Vida+2014}, and \citet[][]{Olah+2009}, respectively. The red squares present data for the M dwarf stars from \citet[][]{Savanov2012}. The gray dots show the data from different surveys presented in \citet[][]{Olah+2009}. The gray line represents the fit to all the data from \citet[][]{Olah+2009} and \citet[][]{Vida+2013,Vida+2014} excluding M stars, while the black and red lines show the fit to the shortest cycles of that dataset. The Sun is shown with its standard symbol. Our measurement is presented with a black square with an error bar. Here, we considered the MAC derived for the flare count.} 
 \label{Fig04}
\end{figure}

To check if the measured magnetic cycle for the KIC\,10544976 secondary star is in agreement with the other low-mass single stars, we use the MAC versus rotation period diagram. Under the reasonable assumption of synchronization between the secondary component and the orbital period of the system, its rotation period was considered to be 0.35~days. The relation between the magnetic cycle and rotation period for the secondary star agrees with the other values measured for low-mass dM single stars (see Figure~\ref{Fig04}). This result strengthens the MAC found in this study. 

Furthermore, it is possible to verify if the secondary star magnetic cycle has some influence at the KIC\,10544976 orbital period by checking the residuals (top panel of Figure~\ref{figo-c}) of the $(O-C)$ diagram. The LS periodogram (bottom panel of Figure~\ref{figo-c}) does not show any significant period around 600 days. Additionally, we can also exclude any variation with a semi-amplitude larger than $\sim$2.6 seconds in the residuals.   

Considering that the LTT effect is the real cause for the KIC\,10544976 orbital period variation, there are two principal scenarios for planetary formation to take place: the first generation of planets formed from a circumbinary protoplanetary disk and the second generation of planets originated from the mass ejected by the common envelope \citep{Perets2011}. \citet{Zorotovic+2013} studied the origin of the ETVs in post-common-envelope binaries and found that the second-generation scenario is more probable than the first-generation one. Due to its magnitude ($R = 18.7$) and sky position, the new generation of giant telescopes, such as the Thirty Meter Telescope\footnote{https://www.tmt.org/} equipped with the chronograph, Planetary System Instrument available at the second generation of instruments, will be crucial to directly confirm the third body around KIC\,10544976.

\section*{ACKNOWLEDGMENTS}
We thank the anonymous reviewer for the careful reading of our manuscript as well as for the useful suggestions to improve it. This study was partially supported by CAPES, FAPESP (L.A.A. and A.D.: 2011/51680-6, LAA: 2012/09716-6, 2013/18245-0, and C.V.R.: 2013/26258-4), and CNPq (C.V.R.: 306701/2015-4). This paper includes data collected by the {\it Kepler} mission. Funding for the {\it Kepler} mission is provided by the NASA Science Mission directorate. This paper makes use of data obtained from the Isaac Newton Group of Telescopes Archive which is maintained as part of the CASU Astronomical Data Centre at the Institute of Astronomy, Cambridge.

\end{document}